\documentclass[sigconf]{acmart}
\AtBeginDocument{%
  \providecommand\BibTeX{{%
    \normalfont B\kern-0.5em{\scshape i\kern-0.25em b}\kern-0.8em\TeX}}}

\copyrightyear{2021}
\acmYear{2021}
\setcopyright{acmcopyright}
\acmConference[CIKM '21] {Proceedings of the 30th ACM International Conference on Information and Knowledge Management}{November 1--5, 2021}{Virtual Event, Australia.}
\acmBooktitle{Proceedings of the 30th ACM Int'l Conf. on Information and Knowledge Management (CIKM '21), November 1--5, 2021, Virtual Event, Australia}
\acmPrice{15.00}
\acmISBN{978-1-4503-8446-9/21/11}
\acmDOI{10.1145/3459637.3482112}

\usepackage{graphicx}
\usepackage{caption}
\usepackage{subcaption}
\usepackage{amsmath}
\usepackage{balance}
\usepackage{scalerel}

\usepackage[binary-units=true]{siunitx}
\DeclareSIUnit{\smallk}{\kilo\relax}
\DeclareSIUnit{\million}{\metre\relax}

\begin{document}

\fancyhead{}

\title{GLocal-K: Global and Local Kernels \\for Recommender Systems}

\author{Soyeon Caren Han}
\authornote{Co-first authors}
\authornote{Corresponding author}
\email{caren.han@sydney.edu.au}
\affiliation{%
  \institution{The University of Sydney}
  \country{Australia}
}

\author{Taejun Lim}
\authornotemark[1]
\email{tlim6535@uni.sydney.edu.au}
\affiliation{%
  \institution{The University of Sydney}
  \country{Australia}
}

\author{Siqu Long}
\email{slon6753@uni.sydney.edu.au}
\affiliation{%
  \institution{The University of Sydney}
  \country{Australia}
}

\author{Bernd Burgstaller}
\email{bburg@cs.yonsei.ac.kr}
\affiliation{%
  \institution{Yonsei University}
  \country{Republic of Korea}
}

\author{Josiah Poon}
\email{josiah.poon@sydney.edu.au}
\affiliation{%
  \institution{The University of Sydney}
  \country{Australia}
}

\renewcommand{\shortauthors}{Han et.al}

\begin{abstract}
Recommender systems typically operate on high-dimensional sparse user-item matrices. Matrix completion is a very challenging task to predict one's interest based on millions of other users having each seen a small subset of thousands of items. We propose a \textbf{G}lobal-\textbf{Local K}ernel-based matrix completion framework, named \textbf{GLocal-K}, that aims to generalise and represent a high-dimensional sparse user-item matrix entry into a low dimensional space with a small number of important features. Our GLocal-K can be divided into two major stages. First, we pre-train an auto encoder with the local kernelised weight matrix, which transforms the data from one space into the feature space by using a 2d-RBF kernel. Then, the pre-trained auto encoder is fine-tuned with the rating matrix, produced by a convolution-based global kernel, which captures the characteristics of each item. We apply our GLocal-K model under the extreme low-resource setting, which includes only a user-item rating matrix, with no side information. Our model outperforms the state-of-the-art baselines on three collaborative filtering benchmarks: ML-100K, ML-1M, and Douban.
\end{abstract}




\maketitle

\section{Introduction}
%background - collaborative filtering background
Collaborative filtering-based recommender systems focus on making a prediction about the interests of a user by collecting preferences from large number of other users. Matrix completion\cite{candes2009exact} is one of the most common formulation, where rows and columns of a matrix represent users and items, respectively. The prediction of users' ratings in items corresponds to the completion of the missing entries in a high-dimensional user-item rating matrix. In practice, the matrix used for collaborative filtering is extremely sparse since it has ratings for only a limited number of user-item pairs.

Traditional recommender systems focus on generalising sparsely observed matrix entries to a low dimensional feature space by using an autoencoder(AE)\cite{zhang2020survey}. AEs would help the system better understand users and items by learning the non-linear user-item relationship efficiently, and encoding complex abstractions into data representations. I-AutoRec\cite{sedhain2015autorec} designed an item-based AE, which takes high-dimensional matrix entries, projects them into a low-dimensional latent hidden space, and then reconstructs the entries in the output space to predict missing ratings. SparseFC\cite{muller2018kernelized} employs an AE whose weight matrices were sparsified using finite support kernels. Inspired by this, GC-MC\cite{berg2018graph} proposed a graph-based AE framework for matrix completion, which produces latent features of user and item nodes through a form of message passing on the bipartite interaction graph. These latent user and item representations are used to reconstruct the rating links via a bilinear decoder. Such link prediction with a bipartite graph extends the model with structural and external side information. Recent studies \cite{rashed2019attribute, strahl2020scalable, ugla2020interpretable} focused on utilising side information, such as opinion information or attributes of users. However, in most real-world settings (e.g., platforms and websites), there is no (or insufficient) side information available about users.

%research aim
Instead of considering side information, we focus on improving the feature extraction performance for a high-dimensional user-item rating matrix into a low-dimensional latent feature space. In this research, we apply two types of kernels that have strong ability in feature extraction. The first kernel, named ``local kernel'', is known to give optimal separating surfaces by its ability to perform the data transformation from high-dimensional space, and widely used with support vector machines(SVMs). The second kernel, named ``global kernel'' is from convolutional neural network(CNN) architectures. The more kernel with deeper depth, the higher their feature extraction ability. Integrating these two kernels to have best of both worlds successfully extract the low-dimensional feature space.

With this in mind, we propose a \textbf{G}lobal-\textbf{Local K}ernel-based matrix completion framework, called \textbf{GLocal-K}, which includes two stages: 1) pre-training the auto-encoder using a local kernelised weight matrix, and 2) fine-tuning with the global kernel-based rating matrix. Note that our evaluation is under an extreme setting where no side information is available, like most real-world cases. The main research contributions are summarised as follows: (1) We introduce a global and local kernel-based auto encoder model, which mainly pays attention to extract the latent features of users and items. (2) We propose a new way to integrate pre-training and fine-tuning tasks for the recommender system. (3) Without using any extra information, our \textbf{GLocal-K} achieves the smallest RMSEs on three widely-used benchmarks, even beating models augmented by side information.

\begin{figure*}[t]
\centering
\includegraphics[width=1.0\textwidth]{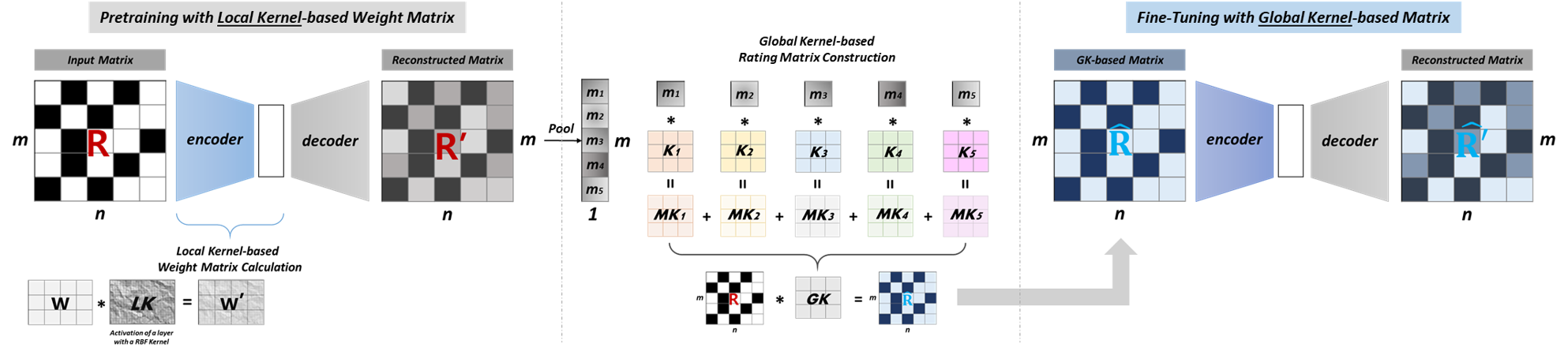}
\vspace{-0.75cm}
\caption{The GLocal-K architecture for matrix completion. (1) We pre-train the AE with the local kernelised weight matrix. (2) Then, fine-tune the trained AE with the global kernel-based matrix. The fine-tuned AE produces the matrix completion result.}
\label{fig:overview}
\end{figure*}

\section{GLocal-K}
Figure \ref{fig:overview} depicts the architecture of our proposed \textbf{GLocal-K} model, which applies two types of kernels in two stages respectively: pre-training (with the local kernelised weight matrix) and fine-tuning (with the global-kernel based matrix)\footnote{The idea of our pre-training and fine-tuning is different from transfer learning.}. Note that we pre-train our model to make dense connections denser and sparse connections sparser using a finite support kernel, and fine-tune with the rating matrix. This matrix is produced from a convolution kernel by reducing the data dimension and producing a less redundant but small number of important feature sets. In this research, we mainly focus on a matrix completion task, which is conducted on a rating matrix $R \in \mathbb{R}^{m \times n}$ with $m$ items and $n$ users. Each item $i \in I =\{1, 2, ..., m\}$ is represented by a vector $r_i = (R_{i1}, R_{i2}, ..., R_{in}) \in \mathbb{R}^{n}$. 
\setlength{\belowdisplayskip}{5pt}
\setlength{\abovedisplayskip}{5pt}

\subsection{Pre-training with Local Kernel}
\textbf{Auto-Encoder Pre-training}\\
We first deploy and train an item-based AE, inspired by \cite{sedhain2015autorec}, which takes each item vector $r_i$ as input, and outputs the reconstructed vector ${r}'_i$ to predict the missing ratings. The model is represented as follows:
\setlength{\belowdisplayskip}{4pt}
\setlength{\abovedisplayskip}{4pt}
\begin{equation}
\label{auto_encoder}
{r}'_{i}=f(W^{(d)}\cdot g(W^{(e)} \ r_{i}+b)+{b}'),
\end{equation}
where $W^{(e)} \in \mathbb{R}^{h \times m}$ and $W^{(d)} \in \mathbb{R}^{m \times h}$ are weight matrices, $b \in \mathbb{R}^{h}$ and ${b}' \in \mathbb{R}^{m}$ are bias vectors, and $f(\cdot)$ and $g(\cdot)$ are non-linear activation functions. The AE deploys an auto-associative neural network with a single $h$-dimensional hidden layer. In order to emphasise the dense and sparse connection, we reparameterise weight matrices in the AE with a radial-basis-function(RBF) kernel, which is known as \textit{Kernel Trick}\cite{giannakopoulos2008novel}.

\medskip

\noindent\textbf{Local Kernelised Weight Matrix}\\
The weight matrices $W^{(e)}$ and $W^{(d)}$ in Eq. (\ref{auto_encoder}) are reparameterised by a 2d-RBF kernel, named \textit{local kernelised weight matrix}. The RBF kernel can be defined as follows:
\setlength{\belowdisplayskip}{4pt}
\setlength{\abovedisplayskip}{4pt}
\begin{equation}
\label{local_kernel}
K_{ij}(U, V) = \text{max}(0, \ 1 - \left \| u_{i}- v_{j} \right \|_{2}^{2}),
\end{equation}
where $K(\cdot)$ is a RBF kernel function, which computes the similarity between two sets of vectors $U$, $V$. Here, $u_i \in U$ and $v_j \in V$. The kernel function can represent the output as a kernel matrix \textbf{\textit{LK}} (see Figure \ref{fig:overview}), in which each element maps to 1 for identical vectors and approaches 0 for very distant vectors between $u_i$ and $v_j$. Then, we compute a local kernelised weight matrix as follows:
\setlength{\belowdisplayskip}{4pt}
\setlength{\abovedisplayskip}{4pt}
\begin{equation}
\label{local_kernelised_weight_matrix}
{W}'_{ij} = W_{ij} \cdot K_{ij}(U, V),
\end{equation}
where ${W}'$ is computed by the Hadamard-product of weight and kernel matrices to obtain a sparsified weight matrix. The distance between each vector of $U$ and $V$ determines the connection of neurons in neural networks, and the degree of sparsity is dynamically varied as vectors are being changed at each step of training. As a result, applying the kernel trick to weight matrices enables regularising weight matrices and learning generalisable representations.

\subsection{Fine-tuning with Global Kernel}
\textbf{Global kernel-based Rating Matrix}\\
We fine-tune the pre-trained auto encoder with the rating matrix, produced by the global convolutional kernel. Prior to fine-tuning, we firstly describe how the global kernel is constructed and applied to build the global kernel-based rating matrix. The entire construction procedure can be defined as follows:
\begin{equation}
    \label{average_pooling}
    \mu_i=\text{avgpool}({r}'_i)
\end{equation}
\begin{equation}
    \label{k_aggregation}
    GK=\sum_{i=1}^{m}\mu_{i}\cdot k_{i}
\end{equation}
\begin{equation}
    \label{convolution}
    \hat{R}=R\otimes GK
\end{equation}
As shown in Figure \ref{fig:overview}, the decoder output of the pre-trained model is the matrix that includes initial predicted ratings in the missing entries, and passed to pooling. With item-based average pooling, we summarise each item information in the rating matrix. Eq. (\ref{average_pooling}) shows the reconstructed item vector $\hat{r}_i$ from the decoder output matrix $R'$ is passed to pooling, and interpreted as item-based summarisation. Let $M=\{\mu_1, \mu_2, ..., \mu_m\} \in \mathbb{R}^{m}$ be the pooling result, which plays a role as the weights of multiple kernels $K = \{k_1, k_2, ..., k_m\} \in \mathbb{R}^{m \times t^2}$. In Eq. (\ref{k_aggregation}), these kernels are aggregated by using an inner product. The result can be dynamically determined by different weights and different rating matrices so that it can be regarded as the rating-dependent mechanism. Then, the aggregated kernel $GK \in \mathbb{R}^{t\times t}$ is used as a global convolution kernel. We apply a global kernel-based convolution operation to the user-item rating matrix for global kernel-based feature extraction. In Eq. (\ref{convolution}), $\hat{R}$ is the global kernel-based rating matrix, which is used as input for fine-tuning, and $\otimes$ denotes a convolution operation.

\medskip
\noindent\textbf{Auto-Encoder Fine-tuning}\\
We then explore how the fine-tuning process works. The global kernel-based rating matrix $\hat{R}$ is used as input for fine-tuning. It takes weights of a pre-trained AE model and makes an adjustment of the model based on the global kernel-based rating matrix, as depicted in Figure \ref{fig:overview}. The reconstructed result from the fine-tuned AE corresponds to the final predicted ratings for matrix completion in recommender system.

\section{EXPERIMENTS}
\subsection{Datasets}
We conduct experiments on three widely used matrix completion benchmark datasets:  MovieLens-100K (ML-100K), MovieLens-1M (ML-1M) and Douban (density 0.0630 / 0.0447 / 0.0152). These datasets comprise of (\SI{100}{\smallk} / \SI{1}{\million} / \SI{136}{\smallk}) ratings of (1,682 / 3,706 / 3,000) movies by (943 / 6,040 / 3,000) users on a scale of \(r \in \{1, 2, 3, 4, 5\}\). For ML-100K, we use the canonical u1.base/u1.test train/test split. For ML-1M, we randomly split into 90:10 train/test sets. For Douban, we use the preprocessed subsets and splits provided by \citet{monti2017geometric}.

%\begin{table}[h]
%\vspace{-0.1cm}
%\caption{Summary statistics of datasets.}
%\vspace{-0.35cm}
%\label{tab:Dataset}
%{\small
%\begin{tabular}{lrrrr}
%\hline
%\textbf{Dataset} & \multicolumn{1}{c}{\textbf{Users}} & \multicolumn{1}{c}{\textbf{Items}} & \multicolumn{1}{c}{\textbf{Ratings}} & \multicolumn{1}{c}{\textbf{Density}} \\ \hline
%ML-100K & 943 & 1,682 & 100,000 & 0.0630 \\
%ML-1M & 6,040 & 3,706 & 1,000,209 & 0.0447 \\
%Douban & 3,000 & 3,000 & 136,891 & 0.0152 \\ \hline
%\end{tabular}
%}
%\vspace{-0.4cm}
%\end{table}

\subsection{Baselines}
We compare the RMSE with the eleven recommendation baselines: \textbf{(1) LLORMA\cite{lee2016llorma} } is a matrix factorization model using local low rank sub-matrices factorization. \textbf{(2) I-AutoRec\cite{sedhain2015autorec}} is a auto-encoder based model considering only the user or item embeddings in the encoder. \textbf{(3) CF-NADE\cite{zheng2016neural}} replaces the role of the restricted Boltzmann machine (RBM) with the neural auto-regressive distribution estimator (NADE) for rating reconstruction. \textbf{(4) GC-MC\cite{berg2018graph}} is a graph-based AE framework that applies GNN on the bipartite interaction graph for rating link reconstruction. We consider GC-MC with side information as \textbf{(5) GC-MC+Extra}. \textbf{(6) GraphRec\cite{rashed2019attribute}} is a matrix factorization utilizing graph-based features from the bipartite interaction graph. We consider GraphRec with side information as \textbf{(7) GraphRec+Extra}. \textbf{(8) GRAEM\cite{strahl2020scalable}} formulates a probabilistic generative model and uses expectation maximization to extend graph-regularised alternating least squares based on additional side information (SI) graphs. \textbf{(9) SparseFC\cite{muller2018kernelized}} is a neural network in which weight matrices are reparameterised in terms of low-dimensional vectors, interacting through finite support kernel functions. This is technically equivalent to the local kernel of GLocal-K. \textbf{(10) IGMC\cite{zhang2019inductive}} is similar to GCMC but applies a graph-level GNN to the enclosing one-hot subgraph and maps a subgraph to the rating in an inductive manner. \textbf{(11) MG-GAT\cite{ugla2020interpretable}} uses attention mechanism to dynamically aggregate neighbor information of each user (item) for learning latent user/item representations.

\subsection{Experimental Setup}
We use two 500-dimensional hidden layers for AE and 5-dimensional vectors $u_i$, $v_j$ for the RBF kernel. For fine-tuning, we use a single convolution layer with a 3x3 global convolution kernel. Inspired by \cite{sedhain2015autorec}, we train our model using the L-BFGS-B optimiser to minimise regularised squared errors, where $L_{2}$ regularisation is applied with different penalty parameters $\lambda_2$, $\lambda_s$ for weight and kernel matrices respectively. Based on validation results, we choose the following settings for (ML-100K / ML-1M / Douban). (1) L-BFGS-B: \textit{$maxiter_{p}$} = (5 / 50 / 5), \textit{$maxiter_{f}$} = (5 / 10 / 5)\footnote{\textit{maxiter} is maximum number of iterations (\textit{p}=pre-training, \textit{f}=fine-tuning).}, (2) $L_2$ regularisation: $\lambda_2$ = (20 / 70 / 10), $\lambda_s$ = (.006 / .018 / .022). We repeat each experiment five times and report the average RMSE results.

\section{Results}
\subsection{Overall Performance}
\begin{table}[t]
\caption{RMSE test results on three benchmark datasets. The column \textit{Extra.} represents whether the model utilises any side information. All RMSE results are from the respective papers cited in the first column, and the best results are highlighted in bold.}
\vspace{-0.3cm}
\label{tab:RMSE_results}
{\small
\begin{tabular}{lcccc}
\hline
\textbf{Model} & \textbf{Extra.} & \multicolumn{1}{c}{\textbf{ML-100K}} & \multicolumn{1}{c}{\textbf{ML-1M}} & \multicolumn{1}{c}{\textbf{Douban}} \\ \hline
LLORMA\cite{lee2016llorma} & - & - & 0.833 & - \\
I-AutoRec\cite{sedhain2015autorec} & - & - & 0.831 & - \\
CF-NADE\cite{zheng2016neural} & - & - & 0.829 & - \\
GC-MC\cite{berg2018graph} & - & 0.910 & 0.832 & - \\
GC-MC+Extra.\cite{berg2018graph} & O & 0.905 & - & 0.734 \\
GraphRec\cite{rashed2019attribute} & - & 0.904 & 0.843 & - \\
GraphRec+Extra.\cite{rashed2019attribute} & O & 0.897 & 0.842 & - \\
GRAEM\cite{strahl2020scalable} & O & 0.917 & - & 0.732 \\
SparseFC\cite{muller2018kernelized} & - & 0.895 & 0.824 & 0.730 \\
IGMC\cite{zhang2019inductive} & - & 0.905 & 0.857 & \textbf{0.721} \\
MG-GAT\cite{ugla2020interpretable} & O & \textbf{0.890} & - & 0.727 \\
GLocal-K (ours) & - & \textbf{0.890} & \textbf{0.822} & \textbf{0.721} \\ \hline
\end{tabular}
}
\vspace{-4mm}
\end{table}

We first evaluated our GLocal-K model on ML-100K (u1.base/u1.test split)/-1M datasets and compare with the baseline models. The RMSE test results are provided in Table \ref{tab:RMSE_results}. It can be easily observed from both GC-MC and GraphRec that incorporate side information improves the recommendation performance, e.g., the error rate of GC-MC+Extra. and GraphRec+Extra. reduce by 0.001 and 0.007 respectively on ML-100K via side information inclusion. Similar to GC-MC, IGMC also learns graph-structural relations from the bipartite user-item interaction graph derived from the rating matrix using GNN but outperforms GC-MC+Extra. by focusing on one-hot sub-graphs with inductive matrix completion. GRAEM focuses on additional graph SI and MG-GAT uses auxiliary information to represent user-user and item-item graph relations. Different from those models above, the first three models in the table use only the rating matrix structure and achieve better results on ML-1M. Our proposed GLocal-K also draws on the rating matrix structure and uses no extra information, outperforming all the baseline models above on three datasets, including those with additional side information, which illustrates the efficacy of combining the local-global kernels for recommendation tasks. Moreover, SparseFC also achieves higher accuracy than those baseline models on three datasets except for MG-GAT, showing the benefits of proper kernel-approximations of the weight matrix. Our GLocal-K surpasses SparseFC, further illustrating the effectiveness  of a global kernel that learns to refine and extract the relevant information from the sparse data matrix. 

\subsection{Cold-start Recommendation}
\begin{figure}[t]
\centering
\includegraphics[width=0.46\textwidth]{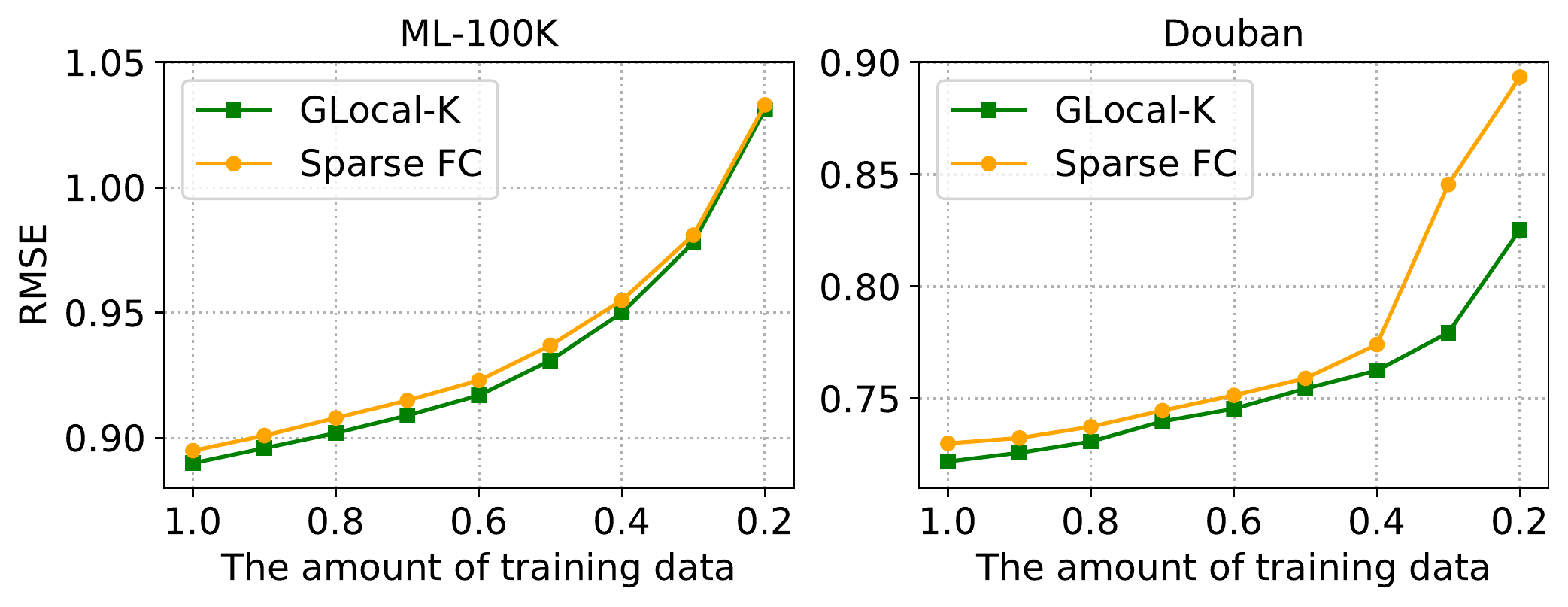}
\vspace{-0.35cm}
\caption{Performance comparison w.r.t.~ different sparsity levels on ML-100K and Douban datasets.}
\label{fig:train_size}
\end{figure}
We varied the training ratio from 0.2 to 1.0 and compared the RMSE test results with SparseFC on ML-100K and Douban in Figure \ref{fig:train_size}. It can be seen that both models on the two datasets demonstrate a similar overall trend: the error rate increases as the training size decreases, which complies with conventional expectation. More specifically, with training ratios of 0.4-1.0, GLocal-K outperforms SparseFC by a merely constant gap on both ML-100K and Douban. This illustrates the superior effectiveness of cooperation by local and global kernels of GLocal-K. In addition, when training size reduces from 0.4 to 0.2 on Douban, the error rate of SparseFC deviates from the previous curve and goes up dramatically while GLocal-K still rises at a stable rate as on ML-100K. This implies that the global kernel can deal with scarce data via feature extraction.

\subsection{Effect of Pre-training}
\begin{figure}[t]
\vspace{-0.2cm}
\centering
\includegraphics[width=0.47\textwidth]{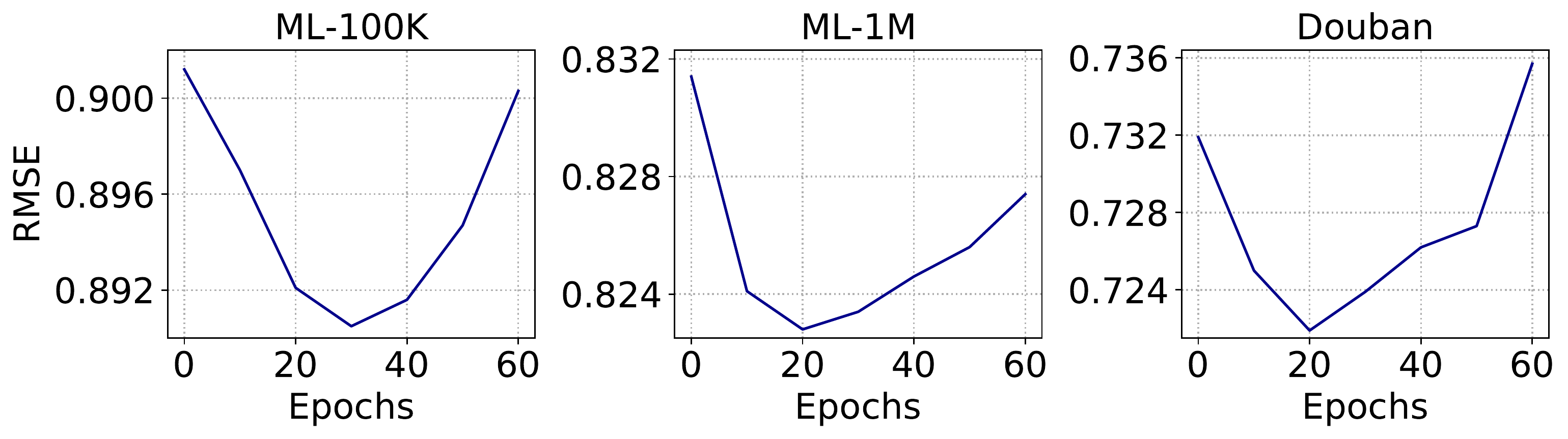}
\vspace{-0.35cm}
\caption{Performance comparison w.r.t.~ the number of pre-training epochs on three benchmark datasets.}
\label{fig:pre_train}
\vspace{-0.4cm}
\end{figure}
We explored the optimal number of epochs for pre-training on ML-100K, ML-1M and Douban. The RMSE results for the three datasets using pre-training epochs from 0 (i.e., no pre-training) to 60 are provided in Figure \ref{fig:pre_train}. These three datasets represent similar bowl-shaped curves. The RMSE first keeps decreasing as the pre-training epoch increases from 0, indicating that pre-training benefits GLocal-K to achieve better performance on all three datasets. Then the RMSE starts to go up again after reaching its optimum at 30 epochs for ML-100K and 20 epochs for both ML-1M and Douban. Referring to the dataset statistics, we surmise that having more item numbers with lower density may lead to less pre-training for optimal performance.

\subsection{Effect of Global Convolution Kernel}
\begin{table}[t]
\caption{Performance comparison of RMSE test results of Global Kernel w.r.t.~ (1) different convolution kernel sizes, (2) different numbers of convolution layers and (3) different kernel aggregation mechanisms on three benchmark datasets. The best results are highlighted in bold.}
\vspace{-0.25cm}
\label{tab:different_parameter}
{\small
\begin{tabular}{cccc}
\hline
 & \textbf{ML-100K} & \textbf{ML-1M} & \textbf{Douban} \\ \hline
\textbf{Kernel size} &  &  &  \\
3x3 & \textbf{0.890} & \textbf{0.822} & \textbf{0.721} \\
5x5 & 0.891 & 0.823 & 0.723 \\
7x7 & 0.891 & 0.823 & 0.723 \\ \hline \hline
\textbf{\# Conv layers} &  &  &  \\
1 & \textbf{0.890} & \textbf{0.822} & \textbf{0.721} \\
2 & 0.893 & 0.827 & 0.725 \\
3 & 0.897 & 0.848 & 0.732 \\ \hline \hline
\textbf{Agg. mechanism} &  &  &  \\
Element-wise & 0.894 & \textbf{0.822} & 0.730 \\
Weighted & \textbf{0.890} & \textbf{0.822} & \textbf{0.721} \\ \hline
\end{tabular}
}
\vspace{-2mm}
\end{table}
To explore the effectiveness of the global kernel-based convolution with in-depth analysis, we first tried multiple kernel sizes and convolution layers. The RMSE results on the three datasets are presented in Table \ref{tab:different_parameter}. It can be seen from Table \ref{tab:different_parameter} that using 3x3 sized kernel achieves the best performance on all three datasets and the error rate goes up as the size increases to 5x5 or 7x7. It implies that focusing on more local features with smaller kernel size might be more effective for extracting generalizable patterns over the whole data matrix. Moreover, Table \ref{tab:different_parameter} shows an incremental performance degradation when the conv layer increases from 1 to 3, indicating a single convolution layer is enough and optimal for feature extraction. In addition, we also explored two variants of kernel aggregation mechanisms: (1) integrating multiple kernels based on the weights and (2) aggregating via pure element-wise average. As shown in Table \ref{tab:different_parameter}, weight-based aggregation reduces RMSE by 0.004 and 0.009 on ML-100K and Douban while achieving similar performance on ML-1M. Overall, it can be seen that using feature-indicative weights to aggregate the kernels is more effective than purely applying element-wise averages.

\section{Conclusion}
In this paper, we introduced \textbf{GLocal-K} for recommender systems, which takes full advantage of both a local kernel at the pre-training stage and a global kernel at the fine-tuning stage for capturing and refining the important characteristic features of the sparse rating matrix under an extremely low resource setting. We demonstrate RMSE on three benchmark datasets: MovieLens-100k/-1M and Douban, outperforming numerous baseline approaches. In particular, we highlighted the effectiveness of our global kernel for exerting scarce data by evaluating the cold-start recommendation. It is hoped that our Global-K gives some insight into the future integration of both kernels for high-dimensional sparse matrix completion with no side information.

%\begin{acks}
%To Robert, for the bagels and explaining CMYK and color spaces.
%\end{acks}

\bibliographystyle{ACM-Reference-Format}
\balance
\bibliography{sample-base}

\end{document}